\documentclass{article}
    \usepackage{csquotes}
    \date{}
    \usepackage[a4paper,top=2cm,bottom=2cm,left=3cm,right=3cm,marginparwidth=1.75cm]{geometry}
    
    \usepackage{graphicx} 
    \usepackage{amsmath}
    \usepackage{amssymb}
    \usepackage{authblk}
    \usepackage{textcomp}
    \usepackage{url}
    \usepackage{verbatim}
    \usepackage{graphicx}
    \usepackage{xcolor}
    \usepackage{pgfplots}
    \pgfplotsset{compat=1.18}
    \usepackage{tabularx}
    \usepackage{siunitx}
    \usepackage{booktabs} 
    \usepackage{subcaption}
    \usepackage{hyperref}
    
\usepackage[
  backend=biber,
  style=numeric-comp,
  maxcitenames=5, 
  maxbibnames=8,
  defernumbers=true, 
  sorting=none       
]{biblatex}
\author[1]{Jan van Delden\thanks{Equal contribution.}\thanks{Corresponding author: jan.vandelden@uni-goettingen.de}}
\author[2]{Julius Schultz\protect\footnotemark[1]} 
\author[2]{Sebastian Rothe}
\author[1]{Christian Libner}
\author[2]{Sabine C. Langer} 
\author[1, 3]{Timo Lüddecke}

\affil[1]{Institute of Informatics, University of Göttingen}
\affil[2]{Institute for Acoustics and Dynamics, Technische Universität Braunschweig}
\affil[3]{Campus Institute Data Science (CIDAS), University of Göttingen}

\title{Minimizing Structural Vibrations via Guided Flow Matching Design Optimization}

\newcommand{\overbar}[1]{\mkern 1.5mu\overline{\mkern-1.5mu#1\mkern-1.5mu}\mkern 1.5mu}
\newcommand{\qoi}{\overbar{L}_v}

\newcommand{\vel}{\boldsymbol{v}}
\newcommand{\sol}{\boldsymbol{u}}
\newcommand{\g}{\boldsymbol{g}}
\newcommand{\p}{\boldsymbol{p}}
\newcommand{\xt}{\boldsymbol{x}_t}
\newcommand{\x}{\boldsymbol{x}}
\newcommand{\shortgrad}{\nabla \mathcal{J}}
\newcommand{\hatgrad}{\hat{\nabla} \mathcal{J}}
\newcommand{\fullgrad}{\nabla_{\boldsymbol{x}_t} \mathcal{J}(r(\boldsymbol{x}_t))}

\begin{document}
\maketitle

\begin{abstract}
Structural vibrations are a source of unwanted noise in engineering systems like cars, trains or airplanes. Minimizing these vibrations is crucial for improving passenger comfort. This work presents a novel design optimization approach based on guided flow matching for reducing vibrations by placing beadings (indentations) in plate-like structures. Our method integrates a generative flow matching model and a surrogate model trained to predict structural vibrations. During the generation process, the flow matching model pushes towards manufacturability while the surrogate model pushes to low-vibration solutions. The flow matching model and its training data implicitly define the design space, enabling a broader exploration of potential solutions as no optimization of manually-defined design parameters is required. We apply our method to a range of differentiable optimization objectives, including direct optimization of specific eigenfrequencies through careful construction of the objective function. Results demonstrate that our method generates diverse and manufacturable plate designs with reduced structural vibrations compared to designs from random search, a criterion-based design heuristic and genetic optimization.
The code and data are available from \url{https://github.com/ecker-lab/Optimizing_Vibrating_Plates/}.
\end{abstract}

\section{Introduction}
\begin{figure}[htb]
\centering
\includegraphics[width=\linewidth]{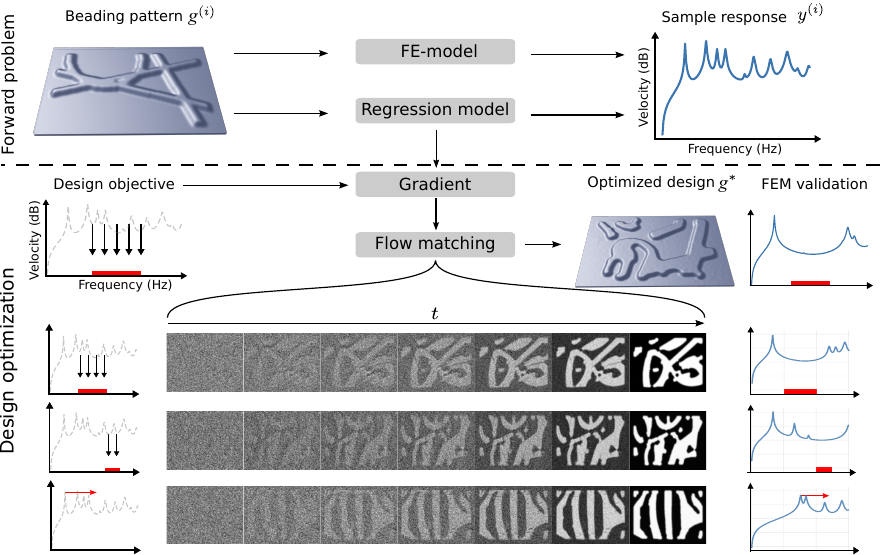}
\caption{Beading patterns on plates change the vibrational behavior depending on the frequency of excitation. We present a deep learning based method to optimize beading patterns to reduce vibrations.}
\label{fig:teaser}
\end{figure}

Unwanted noise emitted by mechanical structures such as cars, trains or airplanes is a common source of discomfort and health issues \cite{basner2014auditory}. The underlying source of noise are structural vibrations caused by mechanical work such as external excitation from an engine.
These vibrations, also called structure-borne sound, are transmitted through the structure and finally radiated in form of air-borne sound to e.g. the inside of a train or car cabin.  
Therefore, noise can be reduced by mitigating structural vibrations, for example by placing beadings.
Beadings are indentations in plate-like structures that locally stiffen the structure and in consequence affect the dynamical behavior of the structure (Figure~\ref{fig:teaser}). They allow to modify the dynamical behavior without changing the overall geometry, which might compromise the primary function of a structure, and without adding or removing material. Building on this principle, our work addresses the design optimization task of placing beadings on fixed-size plates to reduce structural vibrations \cite{delden2024learning}. 
More concretely, we consider structural vibrations in the frequency domain, i.e., the plate is excited by a harmonic loading. Since the system's dynamical response is frequency-dependent, we aim to identify a beading pattern that minimizes vibrations within a user-selected target frequency range. The resulting plate geometry should be manufacturable, it should not exceed material stress limits and should not contain sharp edges or demand excessively high forming ratios.

Vibrational responses of mechanical structures are typically evaluated via numerical simulation with the finite element (FE) method and require finer discretizations with higher excitation frequencies to accurately resolve the occurring wavelength. 
A typical design optimization workflow consists of defining a set of parameters which vary the design, setting up an objective function to optimize, and then performing an optimization over the parameter space based on the objective function, for example with a genetic algorithm. 
This setup requires many evaluations of the objective function depending on the size of the design space. 
Therefore, to keep the optimization computationally feasible, the exploration of the design space needs to be constrained which in turn limits the achievable results. 
This work investigates how the generative deep learning method flow matching can be used for design optimization to address these issues. We apply this method to minimize structural vibrations by optimizing beading patterns in plates. 

\paragraph{Generative deep learning.}
In recent years, generative models have undergone a revolution in effectiveness. The advent of diffusion based methods \cite{sohl2015deep, ho2020denoising, dhariwal2021diffusion} resulted in enhanced generation quality and scalability. Closely related methods based on score matching \cite{song2020score}, and flow matching \cite{lipman2022flow, tong2023improving} have been developed in parallel, leading to improvements in flexibility, quality and efficiency. Recent research highlights the close theoretical and practical similarity between these methods \cite{albergo2023stochastic, ma2024sit}. A common property is that new samples are generated iteratively, by first sampling random noise and then transforming this noise step by step into a new data sample. 
A key question in generative modeling is how to control the generation process. Standard methods include conditional modeling and classifier free guidance \cite{ho2022classifier}, in which the model generates something conditioned on a signal. In consequence, the generative model needs to be trained with pairs of e.g. images and conditional information.
Differently, some methods use two separate models for generation and for controlling the generation. In these methods, the gradient from a model (e.g. a classifier) is used as control information \cite{dhariwal2021diffusion}. Here, gradients have to be obtained based on incompletely generated data points. This can be facilitated by training the guidance model on noisy data points \cite{dhariwal2021diffusion, nichol2021glide}. 
Some works estimate the generated image from noisy images and then use this as input to the guidance model \cite{chung2022diffusion, song2023loss, song2023pseudoinverse}. 
Other works propose to complete the generation task without guidance information and use the final result to acquire gradient information for the generation step \cite{liu2023flowgrad}, which has useful properties but is computationally expensive \cite{wang2024training}.

\paragraph{Generative deep learning for design optimization.}
Generative models have been used for several design optimization tasks, such as structural topology optimization \cite{maze2023diffusion, giannone2023aligning}, ship hull generation \cite{bagazinski2023shipgen}, and airfoil optimization \cite{wu2024compositional, chen2022inverse} and multi-body interaction \cite{wu2024compositional}. 
An important question is how to make the generative model follow constraints on the shape of the generated structure. One approach is the construction of differentiable loss functions that enforce constraints \cite{berzins2024geometry}. Similarly, another work employs a constraint compliance prediction network for guidance \cite{maze2023diffusion}. A recent work casts the sampling process in a diffusion model as a constrained optimization problem and projects the diffusion step onto feasible solutions \cite{christopher2024constrained}.
Closest to this work, a preliminary study addressed the placement of beadings on plates with guided diffusion \cite{delden2024minimizing, delden2024learning}. This work advances this by a focus on manufacturability, employing a more principled guided flow matching approach, and providing extensive experimental validation.

\paragraph{Contribution.}

Our design optimization method follows the framework of classifier guidance: it consists of a surrogate regression model trained to predict vibrational behavior of plates and a generative flow matching model trained to generate novel and manufacturable beading patterns. New beading patterns are generated by the flow matching model through solving an ordinary differential equation (ODE). We augment the ODE velocity model, which defines the ODE's evolution, with guidance information from the regression model. This pushes the generation towards minimizing an objective function designed to reduce structural vibrations.
Specifically, the regression model predicts the dynamical response from a beading pattern. Gradients on the beading pattern pixels are obtained via backpropagation from the objective function on the predicted dynamical response.
The two models in our work have distinct purposes: The flow matching model is trained to generate novel beading patterns following the data distribution in its training dataset. Therefore, the beading patterns in the training data define the design space of the optimization problem. In our case, only beading patterns that fulfill manufacturability conditions are included. The regression model pushes the process towards minimizing the objective function and might cause deviations from the learned path of the flow matching model. To balance this, the relative strength of the guidance from the regression model compared to the flow matching model is carefully controlled.
In summary, our contributions are as follows:

\begin{itemize}
    \item A guided flow matching method to tackle constrained design optimization problems applied for optimizing beading patterns to minimize structural vibrations in plates.
    \item A set of constraints and evaluation to assess the feasibility of generated plate geometries. To fulfill the constraints, our method is designed to balance deviations introduced by guidance.
    \item An evaluation of our method in comparison to genetic optimization, random search and an approach based on a physical heuristic \cite{dissSebastian2022}. Guided flow matching leads to the best result with less compute than genetic optimization.
\end{itemize}

Our approach allows us to combine a regression model with a flow matching model to perform optimization. The flow matching model and it's training data define the design space, which allows us to optimize the beading pattern directly, instead of optimizing design parameters. The following sections first introduce the mechanical background (Section~\ref{sec:physicalmodel}), and our proposed method (Section~\ref{sec:method}). Then, experimental results are presented (Section~\ref{sec:experiments}).

\section{Vibroacoustic benchmark model} \label{sec:physicalmodel}

Vibroacoustic models are crucial to understand the acoustic properties of many engineering systems in the industrial and mobility context. In order to achieve quiet product designs, a good understanding of the sound generation chain is required. This chain includes the excitation or sound source, sound transmission and finally sound radiation. For many vibroacoustic systems the indirect sound transmission path is of main concern: Here, an external excitation induces structure-borne sound energy into the system. This energy then propagates in form of waves through the structure and is radiated in form of air-borne sound to a surrounding fluid. Many engineering systems consist of thin plate-like structures (e.g. chassis of a car, aircraft cabin), which radiate structure-borne sound very efficiently. In consequence, they are prone to generate unwanted noise decreasing user comfort and impacting health. 

Several measures have been proposed in literature to reduce noise emissions of vibroacoustic systems. Primary measures aim to change the properties of the vibrating structure, while secondary measures aim to mitigate already radiated sound (e.g. hearing protection or encapsulation of the structure). In this work, we focus on beading stiffeners, which are primary passive measures to improve the vibroacoustic properties. Passive refers to the fact that the measures do not require external energy supply as opposed to active measures, e.g. active vibration control. 

To evaluate the performance of an applied measure, a common approach is to analyze the vibroacoustic system behavior in the frequency domain. To this end, several assessment criteria have been proposed: \textcite{marburg2001shape} use the first eigenfrequency and aim to maximize this quantity to enlarge the quasi-static regime of the system. \textcite{rothe2019optimization} proposes the mean squared velocity sum level as a suitable target, since it is the driving factor for the radiated sound power. 

Throughout this work, we use a simply supported rectangular plate harmonically excited by a point force as a generic vibroacoustic model. The plate will be modified by beading stiffeners and as a design target we consider the mean squared normal velocity in a target frequency range. The setup of the model is shown in Figure \ref{fig:PlateBeading}a and the detailed geometry, material and damping properties are given in Appendix~\ref{app:platemodel}. A rotational stiffness is included at the boundary to model different boundary conditions, ranging from simply supported (free rotation) to clamped (fixed rotation). Damping is included to the model via a complex Young's modulus $\hat{E} = (1 + i\eta) E$, with constant loss factor $\eta$.

In Section~\ref{sec:beading}, we provide more background on the principle of beading stiffeners, their geometrical properties and how they are manufactured. Based on this information, we present a procedural approach to generate a wide variety of beading patterns and propose a series of simple constraints on the beading geometry to ensure their manufacturability. Section \ref{ssec:FEmodel} then introduces the FE model used to compute the dynamical response of the plate subject to beading stiffeners.
\begin{figure}[b]
\centering
 \begin{tikzpicture}
        \node[anchor=south west] at (0,0) {
            \includegraphics[width=\textwidth]{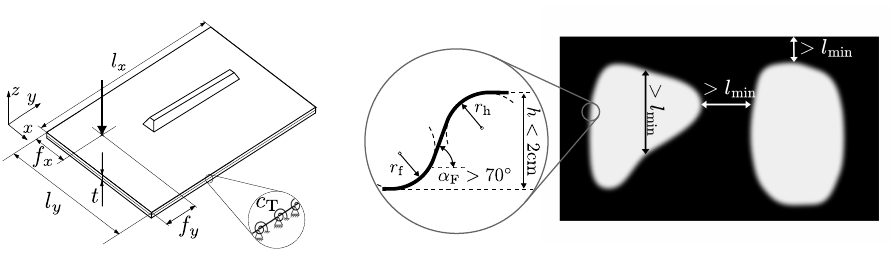}
        };
        \node[anchor=south west, text=black] at (0.01, 0.01) {\small{(\textbf{a})}};
        \node[anchor=south west, text=black] at (0.375\textwidth, 0.01) {\small{(\textbf{b})}};
    \end{tikzpicture}

\caption{(\textbf{a)} Plate model with point force and one exemplary line beading. (\textbf{b}) Design constraint visualized on a beading pattern geometry.}
\label{fig:PlateBeading}
\end{figure}

\subsection{Beading stiffeners} \label{sec:beading}
Beading stiffeners are indentation in a plate geometry with a fixed depth and transition compared to the normal plane of the plate. They are an efficient measure to detune the structural dynamical behavior of a system. Beadings increase the bending moment of resistance without adding additional weight to the system. They are thus well suited for lightweight designs and are incorporated in the structure by gutter-like depressions/elevations perpendicular to the surface.

In this work, we will use a trapezoidal beading cross section as presented in Figure \ref{fig:PlateBeading}b. Trapezoidal beadings are commonly used in industrial applications due to their benefits in terms of manufacturing. In particular, we apply a fixed beading height of $h_{\text{bead}} = \qty{20}{mm}$, a flank angle $\alpha_{\text{F}} = \qty{70}{\degree}$ and the foot and head radii $r_{\text{f}} = r_{\text{h}} = \qty{9.5}{mm}$. This beading cross section is enforced when beadings are incorporated on the plate model as discussed in the next section.

The application of beadings as a measure to detune the structural dynamical behavior has been in the scope of research for decades. Early works as described in \cite{oehler1972steife} mainly focus on an increase in bending stiffness as a design target. In \cite{schriever1994festigkeits} the stiffening effect of beading structures is investigated for car body structures to improve internal and external noise. Design guidelines for acoustic oriented placement of beadings are introduced in \cite{Rieg2018-hy}. These guidelines state, that beading measures should be arranged asymmetrically to avoid repetition of the same geometry in the intermediate area and thus the intermediate areas are not excited by a single eigenfrequency. Moreover, beadings should be directed into stiff structural areas and either tangentially or radially into the force application point. These guidelines provide a good starting point, however, since they aim to be universally applicable, they are not tailored to a specific problem. 

To explore the potential of problem-tailored beading patterns, optimization techniques have been proposed. The work of \cite{Schwarz2002} uses the software OptiStruct, which combines gradient-based optimization with a FE model of a structure subject to beadings. This way, individual load case specific beading patterns are derived for different design targets, namely to maximize the floor and shear stiffness or to maximize the first eigenfrequency. Moreover, in \cite{firl2010optimal}, a gradient-based optimization approach is presented to maximize the first eigenfrequency. The main drawback of such optimization approaches is the computational burden associated with the repeated FE model evaluations. In our work, we do not use gradient-based optimizers as a baseline, as they cannot directly deal with mixed discrete and continuous design parameters. Optimization techniques based on acoustic identification criteria aim to alleviate this computational burden. The goal is to identify a physical criterion, which guides the placement of an optimal beading pattern. Examples for such criteria are the trajectories of the first principal stress \cite{klein1995praxisfahiges}, the bending stress \cite{albers2005new} or the structural intensity \cite{hering2012strukturintensitatsanalyse}. The advantage of such a criterion-based optimization is that it only requires a single model evaluation. We compare our flow matching optimization with an approach, that decides where to place beadings based on the maximum rotational velocity, which we refer to as the rotation-criterion approach \cite{rothe2019optimization, dissSebastian2022}. More details are given in Appendix~\ref{app:critbasedopt}.

\paragraph{Beading pattern variation.} \label{sec:beadinggeneration}

A beading pattern $\g \in \mathbb{R}^d$, represented on a discretized grid with $d$ nodes, defines the specific arrangement of indentations in a plate geometry compared to its normal plane with a fixed height and transition. We represent beading patterns as height maps over the plane of the plate. In the following, they are visualized as monochrome images, where black indicates no indentation and white indicates an indentation (Figure~\ref{fig:FEWorkflow}). To train our generative model, we require a diverse dataset of beading patterns,
To generate diverse beading patterns, we construct a height map generation pipeline by randomly placing geometric primitives such as lines, circles and rectangles on an image. The resulting patterns are randomly mirrored along the x- and y-axes. This procedure is designed to generate a large and diverse design space of beading patterns (Figure~\ref{fig:beading_patterns}).

\paragraph{Design constraints.} 
Beadings in sheet metal components are traditionally produced by deep drawing with specific dies that have to be manufactured individually for each beading pattern. These tools are cost-intensive, which is why they are mainly used in large series production. 
Alternative processes are used to reduce tool costs and increase flexibility. One example is Single Point Incremental Forming (SPIF), which enables the production of complex geometries without special dies through step-by-step forming with a tool head \cite{krasowski2022manufacture}.

During the manufacturing process the plate material undergoes plastic deformation, which may compromise the structural integrity of the plate if the resulting material stress is too high. 
Determining which beading patterns are manufacturable is a complex task \cite{cha2018formability, majic2013development, Schwarz2002}. Therefore, to ensure a degree of practical manufacturability, we impose the following four simplifying constraints on the beading patterns (Figure~\ref{fig:PlateBeading}b):

\begin{itemize}
    \item \textbf{C1}: Minimum distance from the edge = 10\,mm
    \item \textbf{C2}: Bead height: $h_{\text{bead}} \leq 20\,\text{mm}$
    \item \textbf{C3}: Flank angle: $\alpha_F \leq 70^\circ$
    \item \textbf{C4}: Length scale: $l_\text{min} \geq  10\,\text{mm}$
\end{itemize}

\textbf{C1} ensures that beadings are not placed at the boundary of the plate. \textbf{C2} to \textbf{C4} should compromise between the stress load and the stiffness of the beading pattern as proposed by \textcite{Schwarz2002}. \textbf{C4} addresses the minimum length scale, defined as both the minimum width of and minimum space between the beadings. 
Measuring compliance with constraints \textbf{C1} through \textbf{C3} is straightforward, requiring only simple mathematical operations based on their geometric definitions. The measurement is performed on individual pixels of the discretized beading patterns $\g$.
To measure compliance with \textbf{C4}, the minimum length scale of points on the plate has to be checked. To do this, the beading pattern is converted to a binary mask via thresholding to obtain a clear distinction between areas with and without beadings.  A point is considered valid, if it lies inside a circle of diameter $l_{\text{min}}$  that is entirely within areas with or without beadings. Note that the measurement of compliance is performed on discretized beading patterns, which introduces potential for small errors.

\paragraph{Postprocessing.}
During data generation and following design generation, we apply a post-processing procedure to ensure compliance with the constraints. This postprocessing is implemented by image filtering techniques applied to the height map of the plate.
The height map is converted into a binary mask via thresholding. To enforce the minimum length scale constraint (\textbf{C4}), we apply morphological operations, specifically opening and closing, following the approach of \textcite{sigmund2007morphology}. All morphological operations use a circular structuring element with a diameter of $l_\text{min}$. The opening operation, composed of an erosion followed by a dilation, removes beadings that are narrower than the specified scale. Conversely, the closing operation, composed of a dilation followed by an erosion, connects beadings that are too closely spaced. Due to interactions between the opening and closing operations, small violations of \textbf{C4} rarely persist after the postprocessing.
To address the maximum beading angle constraint (\textbf{C3}), the binary mask is convolved with a custom kernel. This kernel is designed to ensure that the cross-sections of the beadings match those illustrated in Figure~\ref{fig:PlateBeading}.

\subsection{Simulation model} \label{ssec:FEmodel}

\begin{figure}[tb]
	\centering
    \includegraphics[width=\textwidth]{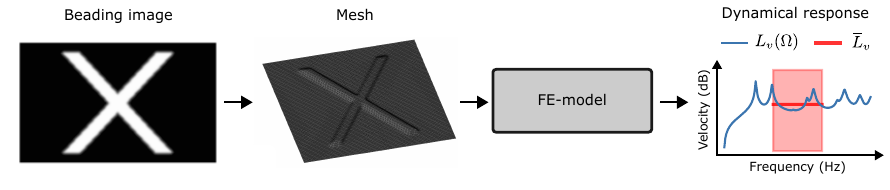}
    \caption{Workflow to incorporate the beading pattern in the FE model: A beading pattern is represented as an depth map, where the values relate to a translation of the mesh nodes in normal direction to the plate surface. The output of the FE model is the mean squared velocity in the frequency domain.}
	\label{fig:FEWorkflow}
\end{figure}
Given a beading pattern $\g$, a vector of model properties $\p$ (boundary conditions, loading) and the excitation frequency $\Omega \in \mathbb{R}^+$, we want to simulate the dynamical response of the system. A common approach to describe the mechanical behavior of arbitrary formed 3D thin structures are shell models. Shell models leverage that one dimension is small compared to the others and represent the geometry on a mid surface. In our case this mid surface will be defined by the plate with beading stiffeners. 
The shell formulation allows to lower the dimension of the mechanical model by integrating over the thickness while still capturing the main mechanical behavior. In particular, a Mindlin-plate formulation is used to represent the bending behavior. This is combined with a disk formulation to represent the in-plane deformations of the mechanical system. 

We use the FE method as a discretization technique to solve the governing system of partial differential equations. To this end, a flat 2D plate is discretized by a regular grid of 121 x 181 nodes and meshed with triangular shell elements using linear ansatz functions. The beading patterns are incorporated in the FE model by shifting the nodes of the mesh in z-direction according to the beading elevation. This process from the beading image to the FE mesh is demonstrated in the left part of Figure \ref{fig:FEWorkflow}.

After discretization and assembling of the system matrices, the discrete dynamical system in the frequency domain is obtained as
\begin{align*}
    \big(-\Omega^2 \mathbf{M} + \mathbf{K}\big) \sol = \boldsymbol{f}.
\end{align*}
Here $\mathbf{M} \in \mathbb{R}^{n \times n}$ and $\mathbf{K} \in \mathbb{C}^{n \times n}$ denote the mass and (complex) stiffness matrix. $\boldsymbol{f} \in \mathbb{R}^{n}$ and $\sol \in \mathbb{C}^{n}$ denote the load and solution vector respectively.

The solution vector $\sol$ contains the solutions of all the nodal degrees of freedom (DOF), which consist of 3 translational and 3 rotational DOF for the applied shell element. For sound radiation the normal displacement and velocity components are of main concern. The normal displacement vector $\sol_{z} \in \mathbb{C}^d$ is a sub-vector of the solution vector $\sol$ and can be obtained by a projection matrix $\sol_{z} = \mathbf{C} \sol$, with $\mathbf{C} \in \mathbb{R}^{d \times n}$. The normal velocity is obtained by differentiating the normal displacement, i.e, $\vel_z= i\Omega \sol_{z}$. The magnitude of this complex quantity is denoted by $\hat{\vel}_z = \vert \vel_z \vert$. The spatially averaged squared velocity is given by
\begin{align}
\label{eqn:spatialavg}
    \overbar{\hat{v}_z^2}(\Omega) = \frac{1}{d} \sum_{j = 1}^d  \hat{v}_{z,j}^2(\Omega).
\end{align}
The mean squared velocity level, also referred to as frequency response function in the following, is then given by 
\begin{align}
\label{eqn:frf_db_scale}
    L_{v}(\Omega) = L_v(\Omega \mid \g, \p) = 10 \log\bigg(\frac{\overbar{\hat{v}_z^2}(\Omega \mid \g, \p)}{v_{\text{ref}}}\bigg)
\end{align}
with the reference velocity $v_{\text{ref}} = \qty{1e-9}{m^2/s^2}$. To emphasize, that the mean squared velocity depends on the beading pattern and the plate parameters, we introduce the notation $L_v(\Omega \mid \g, \p)$.

\section{Optimizing beading patterns with guided flow matching} \label{sec:method}

\begin{figure}[tb]
    \centering
    \includegraphics[width=1\linewidth]{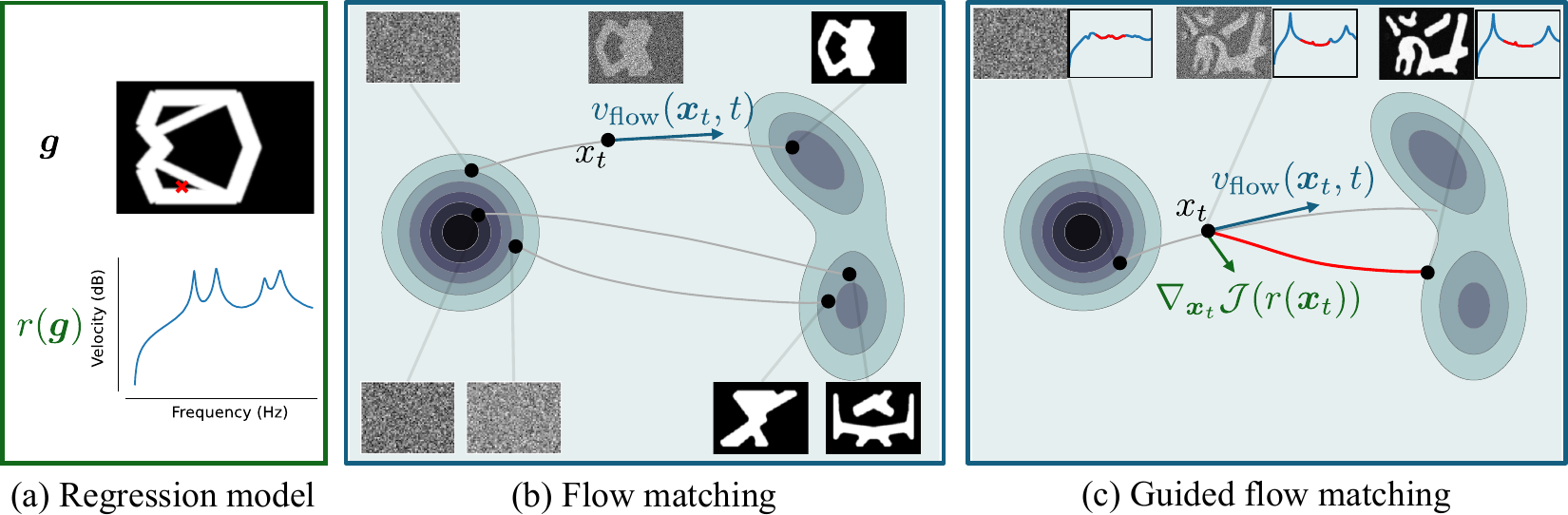}
    \caption{Our method combines a regression model to predict structural vibrations (a) with a flow matching model for manufacturable design generation (b). New plate geometries are generated by solving the differential equation of the flow matching model. An additional step along the gradient of the objective function computed on the prediction of the regression model guides the generation towards minimizing the objective function (c). In (c), the red part of the frequency response (100 Hz - 200 Hz) is minimized.}
    \label{fig:method}
\end{figure}

We address the problem of optimizing the placement of beadings on a thin aluminum plate to reduce structural vibrations. This can be cast as a constrained optimization problem: Our objective is to find a beading pattern geometry $\g$ in the set of all, according to the design constraints, manufacturable geometries $\mathbb{G}$, that minimizes the objective function $\mathcal{J}: C^1(\mathbb{R}) \rightarrow \mathbb{R}$. Here $\mathcal{J}$ takes a frequency response function (Equation \ref{eqn:frf_db_scale}) and maps it to a scalar target quantity. The optimization problem then reads

\begin{equation}
\min_{\g} \, \mathcal{J}\bigl(L_v(\cdot \mid \g, \p)\bigr) \quad \text{subject to} \quad \g \in \mathbb{G}.
\end{equation}

Our method consists of a deep learning based surrogate model $r(\Omega, \g, \p) \approx L_v(\Omega \mid \g, \p)$, trained to predict the frequency response function of the plate given a beading pattern $\g$ and additional properties $\p$ (Figure~\ref{fig:method}a, called regression model), and a generative flow matching model $v_{\text{flow}}$, trained to generate novel beading patterns (Figure~\ref{fig:method}b). In our framework, the role of $r$ is to drive the optimization towards minimizing the objective function and the role of $v_{\text{flow}}$ is to span a design space of manufacturable beading patterns (Figure~\ref{fig:method}c).
In the following, we first separately introduce the flow matching and the regression model as well as their training data, and then describe how the models are combined to arrive at our guided flow matching approach.

\subsection{Flow matching {$v_{\textnormal{flow}}$}}
 
Flow matching \cite{lipman2022flow, lipman2024flow} is a recent generative deep learning method for generating novel data samples from a distribution defined by a training dataset. Flow matching builds upon continuous normalizing flows \cite{chen2018neural}, that are trained to transform samples from a (simple) probability distribution $p$ into another complex probability distribution $q$. This transformation is performed by solving an ordinary differential equation (ODE),

\[
\frac{\mathrm{d}\x}{\mathrm{d}t} = v_{\text{flow}}(\x_t, t)
\]

where $v_{\text{flow}}: \mathbb{R}^d \times [0, 1] \to \mathbb{R}^d$ describes a velocity field. 
Note, this velocity field $v_{\text{flow}}$ is distinct from $\hat{v}$, the physical velocity component of the dynamical response. Instead, it describes the rate of change of samples $\x_t$ from a probability distribution. This system evolves over time $t$, with the state vector $\x \in \mathbb{R}^d$. Here $\x_0$ is set to be a sample from $p$ and $\x_1$ is set to be a sample from $q$. By solving the ODE from 0 to 1 a sample from $p$ can be transformed into a sample from $q$ (Figure~\ref{fig:method}b). For generating novel samples, arbitrary ODE solvers with a flexible amount of steps can be used for a desired accuracy \cite{lipman2024flow}.
To set up a flow matching model, a probability path, i.e., an interpolation between samples of the distributions $p$ and $q$ needs to be defined. The optimal transport interpolant is a popular choice:
\begin{equation}\label{eq:probpath}
\xt = t \, \x_1 + (1-t) \, \x_0
\end{equation}
In this case, the velocity field conditional on the pair $(\x_0, \x_1)$ is constant over $t$, which promotes simple and straightforward to sample ODE paths. During training, samples $\x_0 \sim p = \mathcal{N}(0, \mathbb{I})$, $\x_1 \sim q$ and $t \sim \mathcal{U}[0,1]$ are drawn, which allows for the computation of the ground truth velocity field $v_{\text{gt}} = \x_1 - \x_0$ for $(\x_0, \x_1)$. Then, the MSE between this and the predicted $v_{\text{flow}}(\xt, t)$ is computed as the conditional flow matching loss. Please refer to \textcite{lipman2022flow, lipman2024flow} for a full introduction.

\begin{figure}[tb]
\centering
\begin{subfigure}{0.45\linewidth}
\centering
\includegraphics[width=\linewidth]{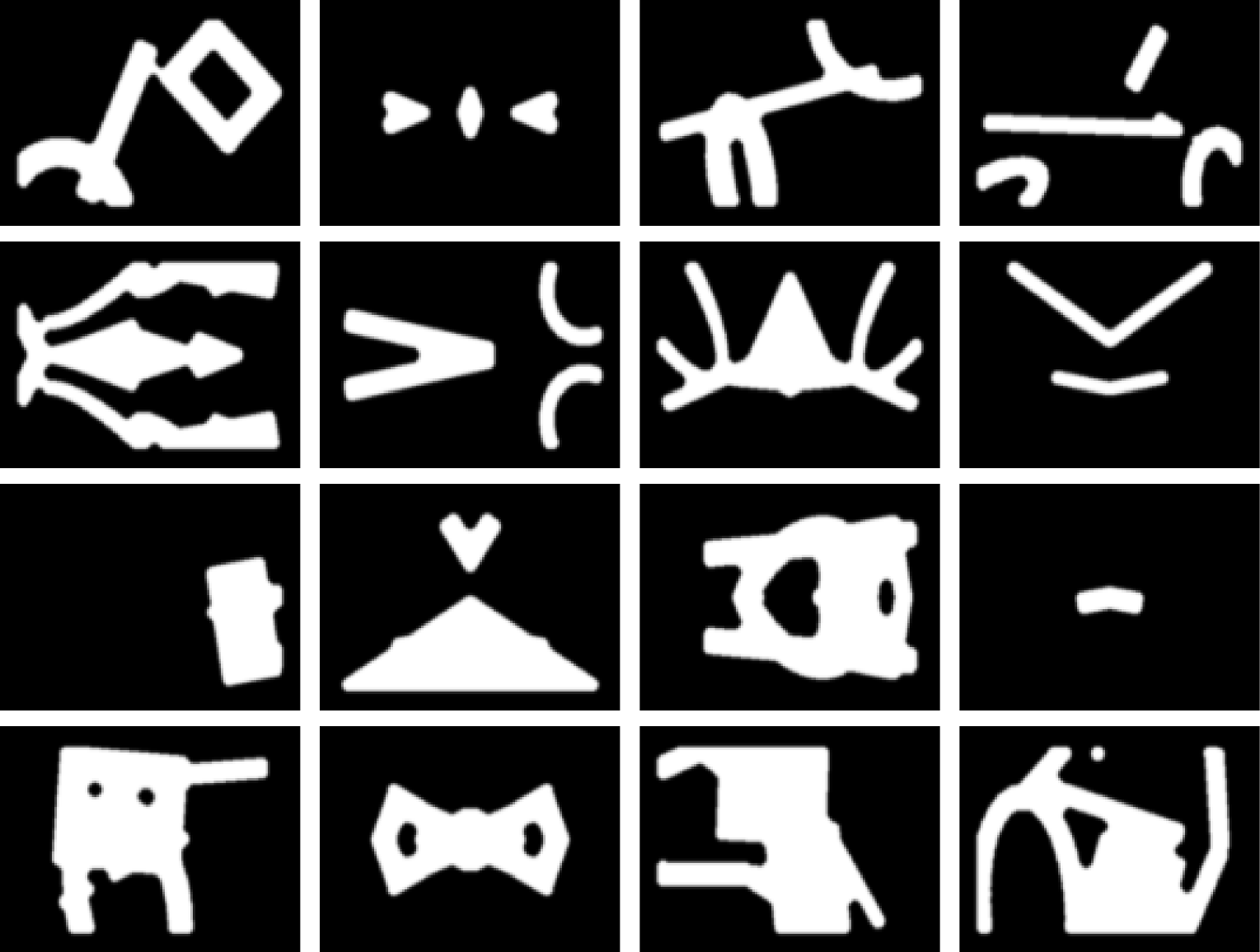}
\caption{Training dataset beading patterns}
\label{fig:beading_patterns}
\end{subfigure}
\hfill
\begin{subfigure}{0.45\linewidth}
\centering
\includegraphics[width=\linewidth]{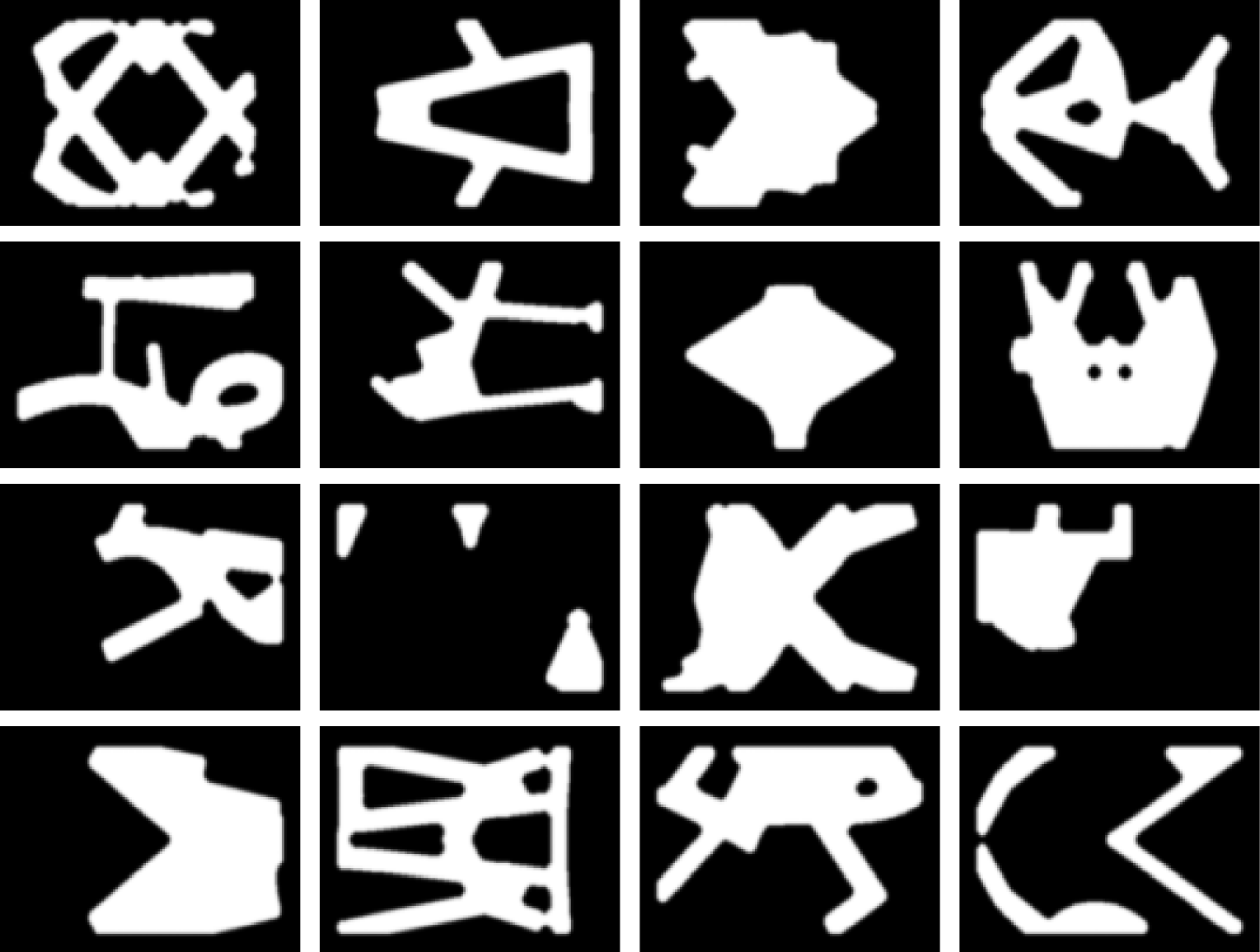}
\caption{Flow matching generated beading patterns}
\end{subfigure}
\caption{Comparison of randomly selected beading pattern samples from the training dataset and beading patterns generated via flow matching without guidance.}
\label{fig:dataset_vs_flowmatching}
\end{figure}

\paragraph{Application to plates.}
We train a flow matching model purely to generate novel beading patterns $\g$. Here, the generated beading pattern $\g$ corresponds to the state $\x_1$ achieved at the end of the generation process ($t$=1). The model is based on the UNet architecture from \textcite{dhariwal2021diffusion}. We perform no conditioning during training of the model, i.e., the model receives no information what kind of beading pattern is supposed to be generated. The model is therefore trained to sample uniformly out of the training data distribution. 
Training is performed on a dataset of \num{300000} random beading patterns, generated as described in Section~\ref{sec:beadinggeneration}. These beading patterns define $q$. The trained model successfully generates novel beading patterns that visually resemble the training data. A minor deviation is that symmetries occur less often in generated beading patterns than in the training data or are sometimes incomplete (Figure~\ref{fig:beading_patterns}).

\subsection{Regression model $r$}

\begin{figure}[tb]
\centering
\includegraphics[width=\linewidth]{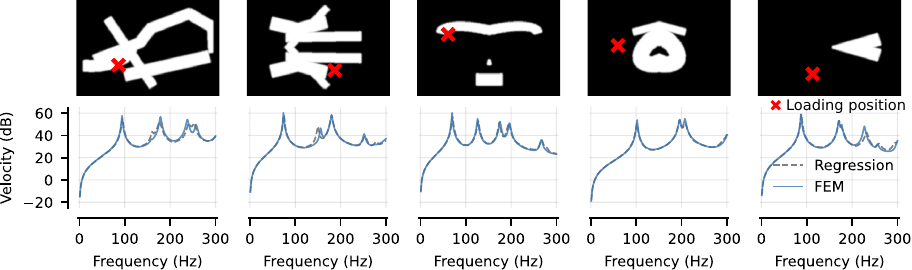}
\caption{Example predictions of regression model $r$ on test dataset with associated FEM ground truth.}
\label{fig:regression_predictions}
\end{figure}

We train a regression model $r$ to predict the velocity field of the vibrations given a frequency $\Omega$, a discretized beading pattern $\g$ and plate properties $\p$, i.e., $r(\g, \p, \Omega)$ is trained to predict $\hat{\vel}_z(\g, \p, \Omega)$, which is directly converted to $L_v(\Omega \mid \g, \p)$ via Equations~\ref{eqn:spatialavg} and \ref{eqn:frf_db_scale}.  Our model is based on the FQO-UNet \cite{delden2024learning} and trained for 400 epochs on the dataset described below. Note that, to acquire predictions for a range of frequencies, the model is evaluated once per frequency. We also denote this as $r(\g)$ for simplicity. Our model is able to accurately predict vibration patterns with minor errors especially for higher frequencies (Figure~\ref{fig:regression_predictions}). 
The regression model is designed to operate on the intermediate states $\xt$ generated by the flow matching model, which represent partially generated beading patterns.
Therefore, random noise was added to $\g$ following Equation~\ref{eq:probpath} during training. This enables the regression model to make predictions for beading patterns that are occluded by different amounts of noise. For this, $t$ was randomly sampled in $[0.25, 1]$, ensuring that the beading pattern is at least partially visible.  No noise was added to the prediction target $\hat{\vel}_z(\g, \p, \Omega)$.

\paragraph{Training data.} \label{sec:training_data}
We generate a dataset of aluminum plates with beading patterns and their respective vibrational response as described in Section~\ref{sec:physicalmodel}. The following properties are varied:
\begin{itemize}
    \item The beading pattern according to the generation procedure described in Section~\ref{sec:beadinggeneration}. 
    Diverse beading patterns improve the generalizability of the trained regression model. Therefore, the constraint (\textbf{C4}) with respect to shape and size of beading patterns is not applied.
    \item The loading position is uniformly varied across the plate surface with a margin of \qty{5}{cm} around the boundary excluded.
    \item The rotational stiffness at the boundary is varied to model different boundary conditions, ranging from simply supported to clamped plates. 
\end{itemize}
The dataset is generated with a specialized FEM software for acoustics \cite{elpaso} and includes \num{50000} different plate samples. Each sample includes a beading pattern $\g^{(i)}$ and plate properties $\p^{(i)}$ along with a vector of excitation frequencies $\boldsymbol{\Omega}^{(i)}$, the corresponding velocity fields $\hat{\vel}_{z}^{(i)}$ and the spatially averaged squared frequency response $\overbar{\hat{\vel}_z^2}^{(i)}$. For each sample, solutions are computed for 15 equidistant randomly shifted frequencies $\Omega$ in the range of 1 - 300 \unit{Hz}. To enhance the available amount of training data, data augmentation was performed by randomly flipping the beading pattern, the loading position, and $\hat{\vel}_{z}^{(i)}$ along the x- and y-axis.

\subsection{Guidance} 

Classifier guidance \cite{dhariwal2021diffusion} is a method developed for diffusion models to steer a generative denoising diffusion model towards generating samples showing a specific type of object. As the name suggests, the idea of classifier guidance is to use a deep learning based classifier $c$ trained to classify the type of object a sample belongs to. As an optimization signal, $\nabla_x \mathcal{J}(c(x))$, the gradient of a loss function on the prediction with respect to the input, is used. 
This quantity can be easily computed via backpropagation, using the automatic differentiation framework of PyTorch \cite{pytorch}. Note, 
backpropagation is conventionally applied to obtain gradients on the weights of a neural network, however obtaining gradients with respect to the input is possible as well.

The method of classifier guidance can be generalized to general optimization targets and can be used in the flow matching framework as well \cite{lipman2024flow}. In our case, we replace the classifier with our regression model, apply it to a beading pattern, compute our objective function on the prediction and perform backpropagation to obtain $\shortgrad  := \fullgrad$. While this quantity could already be directly used for optimization, it does not have any mechanism to promote alignment to the design constraints for the beading patterns, since $\shortgrad$ is defined on individual pixels. 
To incorporate the constraints and geometric priors of our flow matching model, we therefore integrate $\shortgrad$ with $v_{\text{flow}}$ to arrive at a new velocity model: 
\begin{equation} \label{eq:guidance_velocity}
v_{\text{aug}}(\x,t) = v_{\text{flow}}(\x, t) + \alpha\, \beta(t)\, \hatgrad
\end{equation}  
We include a scaling factor $\alpha$ to control the general strength of the guidance term. If not otherwise specified, $\alpha$ is set to 1. We further include $\beta(t)$ defined as a decreasing cosine function:

\begin{equation}
    \beta(t) = \begin{cases} 0.5(1 + \cos(\pi \, t)) (1 - 0.1) + 0.1 & \text{if } t \in [0, 0.75) \\ 0 & \text{if } t \in [0.75, 1] \end{cases}
\end{equation}

In consequence, for the last part of solving the ODE, no guidance is applied and instead pure flow matching is performed. The reasoning behind this design choice is that strong guidance in higher values of $t$ leads to a reduction in the quality of generated beading patterns, since divergence from the flow matching path can not be offset well any more. 
Lastly, we rescale $\shortgrad$:

\begin{equation} \label{eq:gradnorm}
\hatgrad = \frac{\shortgrad \, \|v_\text{flow}\|}{\|\shortgrad \|} 
\end{equation}

This formulation ensures that the gradient component has the same norm as $v_{\text{flow}}$, which allows to directly control the relative influence.

\paragraph{Objective function $\mathcal{J}$.} 
Unless otherwise stated the objective function is the mean squared velocity averaged over a target frequency range:
\begin{align}
    \mathcal{J} := \qoi = \frac{1}{\Omega_2 - \Omega_1} \int_{\Omega_1}^{\Omega_2} L_{v}(\Omega \mid \g, \p) d\Omega. \label{eq:Lbar}
\end{align}
The target frequency range is set to $[100, 200]$~Hz. 
To analyze the efficacy of our method for diverse optimization goals, we further run experiments with the target frequency range $[200, 250]$~Hz and use maximization of the first eigenfrequency as the objective (Appendix~\ref{app:eigenfrequency}).

\subsection{Generating optimized beading patterns}

\begin{figure}[tb]
\centering
\includegraphics[width=\textwidth]{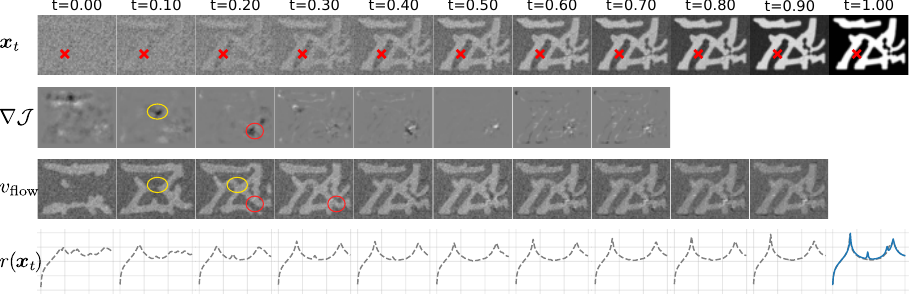}
\caption{The process of guided flow matching to generate a single sample with the Euler method. A subset of the generation steps with $\x_t$, $\shortgrad$, $v_{\text{flow}}$ and $r(\xt)$ are shown in the rows. $\shortgrad$ and $v_{\text{flow}}$ are computed from $\xt$ and combined via Equation~\ref{eq:guidance_velocity} for the ODE velocity model. This is then used to compute the new $\xt$. $\shortgrad$ gradually alters the predicted beading pattern of $v_{\text{flow}}$ leading to an increased complexity of the resulting beading pattern. Negative (black) values in $\shortgrad$ are followed by more beading area in the corresponding area in $v_{\text{flow}}$, as visible in the circled examples.
}

\label{fig:flow_matching_process}
\end{figure}

\begin{figure}[tb]
\centering
\includegraphics[width=5.88in]{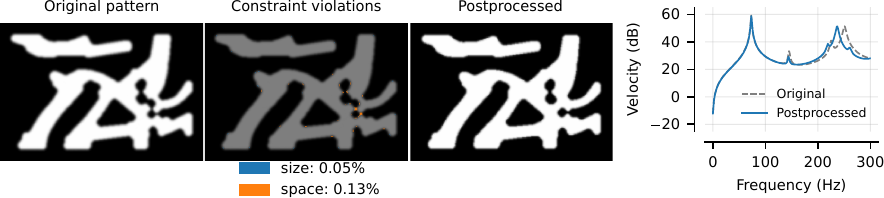}
\caption{We show constraint violations on the final beading pattern from Figure~\ref{fig:flow_matching_process} and the effect of postprocessing. In this example, constraint \textbf{C4} on the minimum space between the beadings is violated in a small area. Violations on the minimum size of beadings occur only for individual pixels and are therefore likely discretization artifacts.}
\label{fig:constraints}
\end{figure}

The flow-matching velocity model $v_{\text{aug}}(\xt,t)$ introduced in the previous section defines an ordinary differential equation (ODE) in $t=[0,1]$. Beading patterns are generated by initializing $\x_0$ from $\mathcal{N}(\boldsymbol{0}, \mathbb{I})$, and then solving the ODE (Figure~\ref{fig:flow_matching_process}). For our experiments, if not otherwise specified, we apply the midpoint method with a fixed step size of 0.05, resulting in a total of 40 steps to generate a new beading pattern. Since the regression model is only evaluated for $t < 0.75$, only 30 steps involve a call to the regression model. The guidance term, $\alpha$ is set to 1. To ensure compliance with our manufacturing criteria, after generating a sample, the postprocessing method described in Section~\ref{sec:beadinggeneration} is applied (Figure~\ref{fig:constraints}). To assess the generation quality, the frequency response of the resulting beading pattern is then obtained via numerical simulation. 

We illustrate the generation by a concrete example (Figure~\ref{fig:flow_matching_process}) such that the interplay and balance of the individual components becomes clearer. Starting from random noise ($\x_0$), in the prediction of the flow matching model $v_{\text{flow}}$ a rough outline of a possible final beading pattern $\x_1$ is already visible and only minor changes in the predicted $\x_1$ over the course of solving the ODE would be expected. However, the guidance gradient $\shortgrad$ leads to deviations from the unguided path to optimize for minimizing the objective function. Negative gradient values (black areas), promote more beading area and seem to often occur between existing beadings in $v_{\text{flow}}$. In consequence, the existing beadings are then connected (red and yellow circled examples in Figure~\ref{fig:flow_matching_process}). In contrast, positive gradient values often split up existing beadings in the prediction. For higher values of $t$, areas in $\shortgrad$ become smaller and would cause more complex structures in the final beading pattern. However, due to the $\beta$ parameter in Equation~\ref{eq:guidance_velocity} this effect is limited. 
Due to the guidance gradient, the process slightly deviates from the learned path of the flow matching model and might therefore also leave the learned distribution of manufacturable beading patterns. Such deviations can result in violations of the manufacturability constraints defined in Section~\ref{sec:beadinggeneration}. In the concrete example, beadings are too close next to each other in a small area (Figure~\ref{fig:constraints}).  The postprocessing filter (Section~\ref{sec:beadinggeneration}) is applied after generation to address constraint violations and involves in this example a minor addition of beading area where violations occurred. Importantly, the optimized frequency response only changes slightly after postprocessing (Figure~\ref{fig:constraints} right).
Quantitatively, generated beading patterns show on average 99.75 \% compliance to the constraints before and 99.89 \% compliance after postprocessing evaluated on 16 generated patterns. 
This compliance percentage (as detailed in Section~\ref{sec:beadinggeneration}) indicates the ratio of pixels without any constraint violations. Beadings in the training dataset have a compliance value of 99.996 \%. Remaining constraint violations are mainly with respect to constraint \textbf{C4} and contain discretization artifacts from single pixels as well as rarely minor actual violations due to interactions in the postprocessing operations.

\section{Experiments} \label{sec:experiments}

In the following, we first compare the performance of several optimization methods with guided flow matching on a simply supported plate (free rotation at the boundary) with the objective to minimize the frequency response in the 100 - 200 Hz range (Section~\ref{sec:methodcomparison}). 
Subsequently, we show how guided flow matching can be used with different optimization objectives, variable load locations and boundary conditions (Section~\ref{sec:differentsettings}). Lastly, we perform experiments to evaluate the effect of different parameters of our method (Section~\ref{sec:ablations}). The specific configurations of the mechanical model are provided in Appendix~\ref{app:plateconfigs}.

For guided flow matching, the generation varies depending on the initial, randomly sampled, $\x_0$. To account for stochasticity, we repeat the generation procedure $n$ times with different random $\x_0$.
Among the generated beading patterns, we select the $k=4$ patterns with the best objective function value according to the regression model $r$ and numerical simulation is performed to obtain validated optimization results. We report the minimum value from the $k$ beading patterns as a final result. 
For our baseline methods, genetic optimization and random search, the $k$ best beading patterns are also selected based on the regression model $r$, evaluated with numerical simulation, and the best result is reported. 
In the following, we set $k$ to four and vary $n$ based on the experiment.

\subsection{Method comparison} \label{sec:methodcomparison}

\begin{figure}[tb]
\centering
\includegraphics{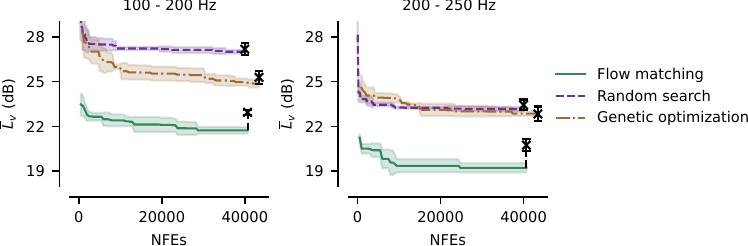}
\caption{Comparison of different optimization methods, given the time and number of neural function evaluations (NFEs) of the regression model allocated to optimization. The curves show $\qoi$ (mean squared velocity level) estimated by the regression model. The black cross shows the final result, evaluated by FEM.}
\label{fig:timeseries}
\end{figure}

\begin{table}[htb]
\caption{Comparison of different optimization methods to minimize structural vibrations. Results for minimizing $\qoi$ (mean squared velocity level) in 100 - 200 Hz and in 200 - 250 Hz are shown.  We report mean (standard deviation) results from four separate runs. Time specified for 200 - 250 Hz objective on a single A100 GPU. Best values highlighted in bold font.}
\label{tab:methodcomparison}
\centering
\footnotesize
\begin{tabular}{lccccc}
\toprule
Optimization range & {100 - 200 Hz} & {200 - 250 Hz}\\
Method & $\qoi$ ($\sigma$) & $\qoi$ ($\sigma$) & NFEs & \# generated plates & Time (\si{min})\\
\midrule
Rotation criterion & 35.2 (-) & 40.6 (-) & - & - & -\\
Random search & 27.2 (0.4) &  23.4 (0.6) & \num{40000} & \num{40000} & \textbf{9} \\ 
Genetic optimization  & 25.3 (0.4) & 22.8 (0.4) & \num{43430} & \num{43430} & 27 \\ 
\textbf{Flow matching} & \textbf{22.9} (0.1) & \textbf{20.7} (0.4) & \num{40672} & \num{1312} & 15 \\ 
\bottomrule \\
\end{tabular}
\end{table}

We compare our guided flow matching method to three baseline methods: 
\begin{enumerate}
\item \textbf{Rotation criterion}, an engineering heuristic based on the rotation velocity of the plate without beadings, which has been identified as a good placement criterion \cite{dissSebastian2022}.
\item \textbf{Random search}, where beading patterns are randomly generated based on the procedure described in~\ref{sec:beadinggeneration} and evaluated with the regression model. 
\item \textbf{Genetic optimization}, where in total 43 parameters describing the beading pattern are optimized. 
\end{enumerate}
Details for these methods are given in Appendix~\ref{app:baselinemethods}. 
The regression model constitutes most of the computational demand, since it needs to be evaluated once per frequency. To conduct a fair comparison, the number of frequency responses evaluated by the regression model (in short NFE for neural function evaluations) is fixed for the methods at around \num{40000} with minor variation due to batch effects.
Consequently, only \num{1312} beading patterns are evaluated for flow matching since multiple NFEs are required for each generation. The optimization based on the rotation criterion does not require any evaluations of the regression model.

\paragraph{Results.}
Guided flow matching leads to the lowest $\qoi$ compared to the baseline methods for a simply supported plate (Table~\ref{tab:methodcomparison}), with an additional reduction of \qty{9}{\percent} ($2.4$~dB) for the 100 - 200 Hz optimization range and \qty{9}{\percent} ($2.1$~dB) for the 200 - 250 Hz optimization range compared to the second best method.
We also compare how the best result, as evaluated by the regression model, changes over the runtime of an optimization (Figure~\ref{fig:timeseries}). Our approach already achieves a competitive result, better than the final result of the baseline methods at the beginning of the optimization after a few seconds. This highlights the efficiency of our method. 
An issue with gradient-based optimization based on learned regression models is that the process might exploit weaknesses in the regression model, since the optimized beading pattern might differ from the training data distribution. In consequence, the actual result validated by numerical simulation could be much worse than the result as predicted by the regression model. This effect is partially visible in our results (Figure~\ref{fig:timeseries}, black x compared to green curves), as there is an average difference of $1.5$~dB between prediction and numerical simulation in the final result. This effect is inhibited by the flow matching model, which pushes the optimization towards the learned design space. 
Overall, guided flow matching presents clear advantages over the compared baseline methods, delivering better final performance with greater efficiency.

\begin{figure}[tb]
\centering
\includegraphics[width=\textwidth]{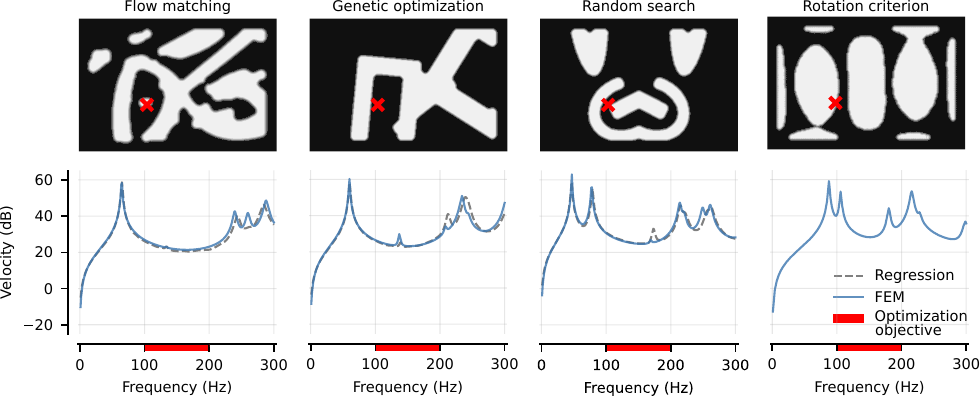}
\caption{Minimizing response in 100 - 200 Hz. We compare the performance of diverse optimization methods for simply supported plates. Loading position marked by red x.}
\label{fig:qualitative_results_methods}
\end{figure}

Beading patterns generated by our guided flow matching method are more complex in comparison to the other methods (Figure~\ref{fig:qualitative_results_methods}, additional visualizations in Appendix~\ref{app:visualizations}). While random search and genetic optimization are strictly constrained to the parametric design space, beading patterns from guided flow matching can deviate from its training set. This is demonstrated, for example, by the generated plates containing bulges on beadings and changes in width over a beading, that were not part of the original design space. 
Generated beading patterns for the same objective function from different random initializations differ, but exhibit common properties such as the enclosing of the loading position, with an opening to the lower edge of the plate (Figure~\ref{fig:qualitative_results_methods} flow matching and genetic optimization, and Appendix~\ref{app:visualizations}).

\subsection{Variation of mechanical model and objective function} \label{sec:differentsettings}

\begin{figure}[tb]
\centering
\includegraphics[width=\linewidth]{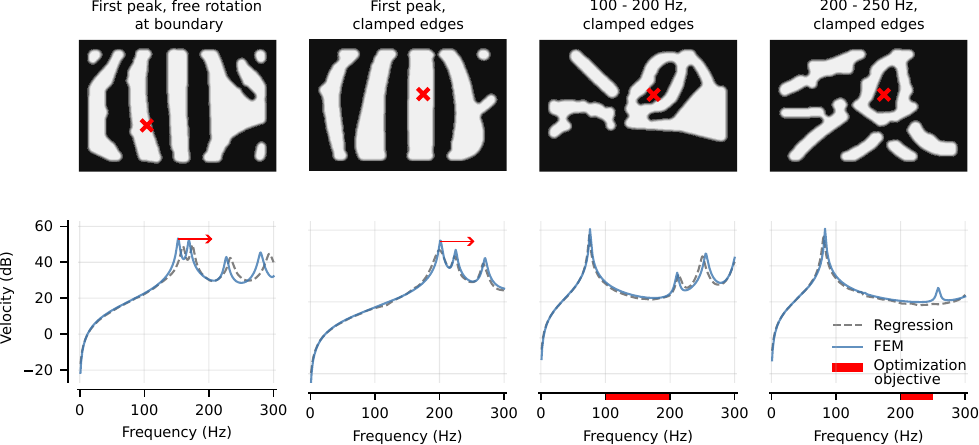}
\caption{Optimized beading patterns differ depending on the objective function and the boundary condition of the plate model. Objective function and setting is stated above the beading patterns.}
\label{fig:different_configs}
\end{figure}

We apply our method to diverse boundary conditions, loading positions, target frequency ranges, and an objective function constructed to maximize the first eigenfrequency (Figure~\ref{fig:different_configs}).
%
When maximizing the first eigenfrequency, the resulting beading patterns contain four thick vertical beadings with some space between each other (Figure~\ref{fig:different_configs}, left half). The beading patterns for the clamped plate and simply supported plate look quite similar. Also, a beading is placed directly at the load application point. 
%
The optimization result for minimizing the response in the 100 - \qty{200}{\Hz} range for the clamped plate model (Figure~\ref{fig:different_configs}, mid right) looks distinctly different to the result for the simply supported plate (Figure~\ref{fig:qualitative_results_methods}, left). The total beaded area is smaller and the individual beading elements cover a larger area. The result for minimizing 200 - \qty{250}{\Hz} (Figure~\ref{fig:different_configs}, right) contains more smaller elements in comparison.
Collectively, guided flow matching showcases adaptability, producing distinct beading patterns tailored to diverse objectives and physical conditions.

\subsection{Method component analysis} \label{sec:ablations}

\begin{figure}[tb]
\centering
\begin{subfigure}[t]{0.27\linewidth}
\includegraphics[height=1.30in]{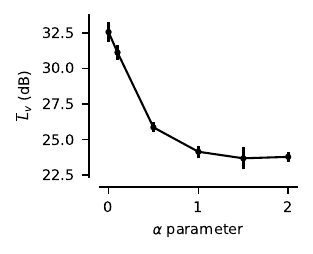}
\caption{\footnotesize\centering $\alpha$ in $v_{\text{aug}}$}
\end{subfigure}
\begin{subfigure}[t]{0.278\linewidth}
\includegraphics[height=1.30in]{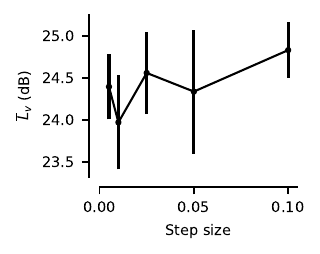}
\caption{\footnotesize\centering Step size in ODE solver}
\end{subfigure}
\begin{subfigure}[t]{0.21\linewidth}
\includegraphics[height=1.3in]{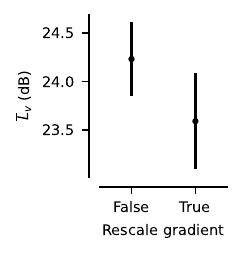}
\caption{\footnotesize\centering Rescaling of $\shortgrad$}
\end{subfigure}
\begin{subfigure}[t]{0.22\linewidth}
\includegraphics[height=1.3in]{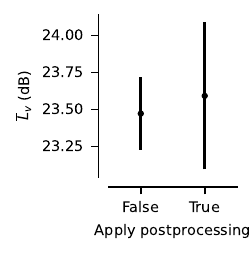}
\caption{\footnotesize\centering Postprocessing beading pattern}
\end{subfigure}
\caption{$\qoi$ on y-axis is given as the mean of the four best plates out of 160 generated plates, with error bars giving the standard deviation. For the ablation of the rescaling of $\shortgrad$, $\alpha$ was set to 5000 to match the total influence of the guidance term with rescaling. Different values for $\alpha$ did not lead to better results.}
\label{fig:alphasweep}
\end{figure}

We evaluate the influence of the $\alpha$ parameter in Equation~\ref{eq:guidance_velocity} (Figure~\ref{fig:alphasweep}a) that controls the strength of the guidance. The mean $\qoi$ improves up to a value of 1.5. Setting $\alpha$ to 1 leads to only a marginally worse $\qoi$, has less variance and represents a simpler choice. Therefore we adopt $\alpha=1$. The step size of the ODE solver does not seem to have a systematic influence on $\qoi$ (Figure~\ref{fig:alphasweep}b). Rescaling $\shortgrad$ to the norm of $v_{\text{flow}}$ following Equation~\ref{eq:gradnorm} has a positive effect compared to scaling it with an absolute value (Figure~\ref{fig:alphasweep}c). Our postprocessing step, which ensures generated plates fulfill manufacturability criteria, did not lead to worse results. A slight increase in variance of $\qoi$ is observed (Figure~\ref{fig:alphasweep}d).

\section{Discussion}

In this work, we presented a design optimization method based on guided flow matching to generate beading patterns on plates that minimize structural vibrations. The resulting beading patterns outperform baseline methods by achieving around \qty{9}{\percent} lower mean vibration values compared to the best baseline. The optimization for 8 beading patterns in parallel takes around 11 seconds on an A100 GPU. 
Our method is adaptable to a broad range of objective functions such as different frequency ranges and variations in the boundary condition and loading position. 
This versatility extends to maximizing the first eigenfrequency, for which we present a differentiable objective function enabling gradient-based optimization.
Generated beading patterns are diverse and exhibit distinct characteristics depending on the mechanical model and objective function. They remain close to the design space of the training data as they consist of the same geometric primitives. However, the generated beadings are less regular, vary in width, and include more individual elements. When targeting the same objective function, generated patterns often vary in some areas, but some general characteristics can be observed. For instance, for some objective functions, the patterns consist of lines with similar orientation. Furthermore, very often a beading either directly passes the force application point or the force application point is surrounded by beadings. This is in accordance with textbook design guidelines for acoustic oriented placement of beadings \cite{Rieg2018-hy}. 

We further emphasized the manufacturing process of beading patterns generated in this work based on simplified manufacturing constraints. The generative flow matching model is trained to follow the distribution of its training data, which adheres to these manufacturing constraints, and is therefore designed to generate manufacturable beading patterns. 
Through this data-driven approach, the generative model learns to incorporate manufacturing constraints implicitly. Consequently, the optimization process itself does not require them to be added as explicit, separate conditions.
In our method, the flow matching process runs in parallel to gradient-based optimization. The gradient term can lead to deviations from the probability path on which the flow matching model was trained. In consequence, generated beading patterns with guidance might also exhibit deviations from the training data distribution. We address these issues with a postprocessing step, that removes most violations of our manufacturability criteria. A promising direction for follow-up work is incorporating manufacturing constraints more directly into the flow matching process, e.g. via differentiable constraint functions \cite{berzins2024geometry} or via constraining the flow matching velocity model \cite{christopher2024constrained}.

In this work, we addressed the design of beading patterns on rectangular plates. Applying and extending our method to different design tasks, such as the placement of damping material, or to more complex mechanical structures like curved shells or multi-component structures, is a challenging and interesting future research direction. Furthermore, exploring multi-disciplinary design, for instance by considering vibroacoustic as well as strength analysis objectives, presents another interesting research direction. 

\paragraph{Limitations.}
Our method entails a high upfront computational investment, primarily for generating the regression model training dataset (e.g., the regression model training dataset involved  \num{50000} plate samples simulated at 15 frequencies) via costly numerical simulations, and subsequently training the deep learning models (flow matching and regression). A breakdown of the total computational cost is given in Appendix~\ref{app:compute}. However, these resources are reusable, for example, for optimization tasks with different boundary conditions or different objective functions. Also, compared to genetic optimization performed directly with numerical simulations for each design evaluation, our approach demands much less compute to achieve an optimized design. 
For instance, a direct simulation-based genetic optimization, with the same parameters as in Table~\ref{tab:methodcomparison}, would require more than five times the numerical simulations used for generating the training dataset. Therefore, while our method requires a high computational cost before optimization, it might still be more efficient than alternatives.
Another limitation is the method reliance on a surrogate model for guidance. While the flow matching component helps maintain design validity, the process can still exploit surrogate inaccuracies, leading to observed discrepancies (around 1.5 dB in our results) between predicted and validated performance.
Third, the generated beading patterns often exhibit greater complexity and less regularity than those from traditional parametric approaches. While they meet the geometric constraints defined in this work, actually manufacturing them might pose more difficulties than simpler patterns.

\paragraph{Author contributions.}
J. van Delden and J. Schultz coordinated the work and conceptualized the experiments. 
J. van Delden developed and conceptualized the guided flow matching method and regression model and performed the experiments and ablations.
J. Schultz implemented the mechanical model, generated the dataset and performed the genetic optimization baseline.
C. Libner contributed the manufacturability criteria and formulation of peak position loss. 
S. Rothe contributed results based on the rotation criterion optimization method. 
All previously mentioned authors contributed to the initial draft.
S. C. Langer and T. Lüddecke contributed equally to supervision and revision of the initial draft.

\paragraph{Acknowledgements.}
This research is funded by the Deutsche Forschungsgemeinschaft (DFG, German Research
Foundation), project number 501927736, within the DFG Priority Programme 2353: Daring More Intelligence - Design Assistants in Mechanics and Dynamics’.
The authors gratefully acknowledge the computing time granted by the Resource Allocation Board and provided on the supercomputer Emmy/Grete at NHR-Nord@Göttingen as part of the NHR infrastructure. The calculations for this research were conducted with computing resources under the project nii00194.

\paragraph{Declaration of generative AI and AI-assisted technologies in the writing process.}
During the preparation of this work the authors used standard large language models in order to paraphrase text and generate basic code. After using this tool/service, the authors reviewed and edited the content as needed and take full responsibility for the content of the published article.

\paragraph{Declaration of competing interests.}
S. Rothe is co-founder of Noise2zero GmbH, a company offering engineering services in acoustics. Otherwise, the authors declare no competing interests.

\printbibliography
\clearpage

\appendix
\section*{Appendix}
\section{Plate model}  \label{app:platemodel}
\subsection{Properties}
\begin{table}[htb]
    \caption{Geometry and material parameters of the plate}
    \label{tab:plate_prop}
    \centering
   \footnotesize
    \begin{tabular}{lccccccc}
    \toprule
\multicolumn{3}{c}{Geometry} & \multicolumn{4}{c}{Material (Aluminum)}  \\
\cmidrule(lr){1-3}
\cmidrule(lr){4-7} 
length & width & thickness &  density & Young's mod. & Poisson ratio & loss factor  \\
\midrule
 $0.9\,$m & $0.6\,$m & $0.003\,$m &  $2700\,$kg/m$^{3}$ & $7\text{e}10\,$N/m$^{2}$ & 0.3 & 0.02 \\
    \bottomrule \\
    \end{tabular}
\end{table}

\subsection{Configurations} \label{app:plateconfigs}
The rotational boundary condition is modeled via a rotational stiffness around the x or y axis for the respective edges. The rotational stiffness is added in form of rotational springs at the respective boundary nodes and the values for the rotational stiffness are given the Table below. 

\begin{table}[htb]
\caption{Loading and boundary condition parameters for the plate configurations.}
\label{tab:plate_prop_load_bc}
\centering
\footnotesize
\begin{tabular}{lccc}
\toprule
& \multicolumn{2}{c}{Loading (Point force)} & Boundary condition (rot. stiffness) \\
\cmidrule(lr){2-3} 
\cmidrule(lr){3-4} 
Setting &x-position & y-position & $c_{ry}$/$c_{rx}$ \\
\midrule
Free rotation & $0.31\,$m & $0.21\,$m & $0.0\,$ Nm/rad\\
Clamped & $0.52\,$m & $0.35\,$m & $100.0\,$ Nm/rad\\
\bottomrule \\
\end{tabular}
\end{table}

\clearpage

\section{Alternative optimization methods} \label{app:baselinemethods}

\subsection{Rotation-criterion based optimization} \label{app:critbasedopt}

For the criterion-based optimization, we use the method proposed by \textcite{dissSebastian2022}. This is a heuristic approach for acoustic-oriented beading design, relying on a one-shot optimization strategy using a single frequency-domain analysis of the unbeaded structure. Unlike traditional optimization techniques, no iterative loops are required. Instead, the plate is solved directly in the frequency domain, and the distribution of a physical quantity within the structure  is computed (in this case, the squared rotational velocity is used). This involves a decision-making process to determine which quantity is most suitable. A detailed discussion of this is beyond the scope of this work and is presented in \textcite{dissSebastian2022}. The physical values are then integrated over a defined target frequency range using a Riemann sum, serving as the identification criterion. Taking into account the C1-C4 constraints:
\begin{itemize}
    \item \textbf{C1}: Minimum distance from the edge = 10\,mm
    \item \textbf{C2}: Bead height = 20\,mm
    \item \textbf{C3}: Flank angle = 70$^\circ$
    \item \textbf{C4}: Length scale $l_\text{min} >$ 10\,mm
\end{itemize}

\noindent bead geometries are modeled into the FE mesh until a beading ratio of 1.0 is achieved (50\% of the plate area is modified). This corresponds to an optimum in terms of bending resistance moment and area moment of inertia. The beading process starts in regions with the highest values of the identification criterion. After modeling the beading pattern, the modified structure is re-simulated. A significant insulating effect within the target frequency range is expected as a result.

\subsection{Genetic optimization} \label{app:geneticopt}
Genetic algorithms are heuristics to solve optimization problems via operations inspired from biology, e.g. selection, cross-over or mutation. These class of algorithms are gradient-free and follow an iterative procedure: The starting point is a population of so called candidate solutions, which are randomly selected in the search space. The fitness of each candidate is determined by evaluating an objective function and the best candidates are randomly selected as parent candidates for the next generation. The new generation of candidate solutions is obtained by recombination of the selected parent candidates and additional mutation operations.
Genetic algorithms can handle large search spaces and are able to escape local minima, however, they suffer from slow convergence rates and no guarantee to find an optimal solution.

To apply the genetic algorithm to our beading pattern design optimization problem, it requires a parametric representation of the design space. Here, we use a parametrization of the geometric primitives (lines, ellipse and rectangles) as introduced in Section~\ref{sec:beadinggeneration} to define the beading pattern dataset. A line is parametrized by 5 parameters (start point, end point and width), an ellipse is parametrized by 8 parameters (length, width, mid point, line width, rotation angle, switch parameter for full or partial ellipse and an angle which defines the angle of an partial ellipse) and the rectangle is parametrized by 7 parameters (length, width, mid point, line width, rotation angle and a switch parameter for full or parially filled rectangle). Additionally 3 hyper-parameters control the numbers of line, ellipse and rectangle objects with a maximum of 2. Thus in total $2*(5+8+7) + 3 = 43$ parameters define a beading pattern. The objective function is $\qoi$ as defined in \ref{eq:Lbar}.   

We use the differential evolution algorithm \cite{storn1997differential} implemented in the scipy optimization package. The default parameters for mutation and recombination are applied as well as a population size of $10$ and a maximum number of iterations of $100$.

\subsection{Random search} \label{app:randomsearch}

To perform random search, the procedural beading pattern generation method described in Section~\ref{sec:beadinggeneration} is used to generate a large amount of beading patterns with random parameters. The frequency response of each of these beading patterns is then evaluated with the regression model. Therefore, the number of evaluated beading patterns is the same as the amount of neural function evaluations. From all evaluated beading patterns, the four with the best objective function value are selected and numerical simulation is performed to obtain ground truth frequency responses. The beading pattern with the best objective function value according to numerical simulation is the final optimization result.

\section{First eigenfrequency objective function} \label{app:eigenfrequency}

Our objective is to shift the first peak of the frequency response function to a higher frequency.
Let $L_v $ be frequency response function in dB scale, as defined in Equation (\ref{eqn:frf_db_scale}). With scipy's find\_peaks we can find the frequencies of all peaks $\Omega_{p1} , \Omega_{p2} ,..., \Omega_{pN} $.  The find\_peaks operation is not differentiable with respect to the FRF. Therefore, we define our custom objective function:

\begin{equation}
   \mathcal{J} := - \frac{\int_0^{\Omega_p}  \Omega e^{\beta 
   L_{v}(\Omega \mid \g, \p)
   } d\Omega}
    {\int_0^{\Omega_p} e^{\beta 
    L_{v}(\Omega \mid \g, \p)
    } d\Omega}
    \label{eq:first_peak_position_loss}
\end{equation}

where $\Omega_p = \frac{1}{2}(\Omega_{p1}+\Omega_{p2})$. 
Formula \eqref{eq:first_peak_position_loss} can be interpreted, as converting the frequency response function $L_v$ to a probability distribution via a softmax and taking the mean of this distribution. Since the first peak $\Omega_{p1}$ is the maximum of $\tilde{L}_v$ in the  interval $[0,\Omega_p]$, for $\beta \rightarrow \infty$ the objective function $J_{FPP}$ approaches $-\Omega_{p1}$. However, we have found that for $\beta=1$ the objective function $\mathcal{J}$ approximates the first peak frequency reasonably well. During the flow matching procedure, the predicted frequency response function can be noisy. This leads to wrongly detected peaks by the find\_peaks operation. To overcome this issue, we smooth the predicted frequency response function with a Gaussian blur. 

\section{Computational resources} \label{app:compute}
All computations were performed within the NHR-Nord cluster environment. Neural network related computation where employed on a system with a single A100 GPU. Training of the flow matching model took around 24 h, training of the regression model took around 19~h. Performing the numerical simulations for the \num{50000} beading patterns in the training dataset with 15 frequencies per plate, took in total around 250~hours utilizing 15 cores in parallel. 

\clearpage
\section{Additional visualizations} \label{app:visualizations}

\begin{figure}[htb]
\centering
\vspace{-0.3in}
\includegraphics[width=\textwidth]{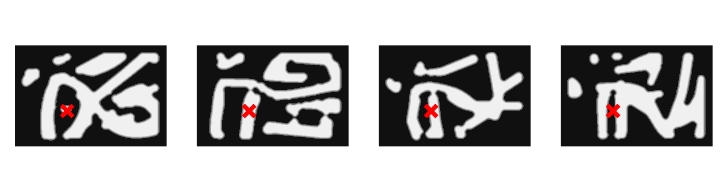}
\includegraphics[width=5.38in]{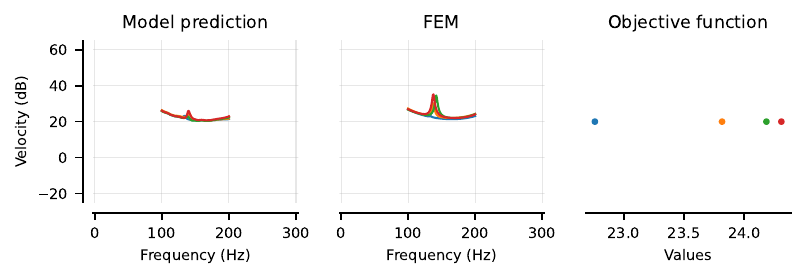}
\caption{Minimizing the response in 100 - 200 Hz. Plate with free rotation at boundary. Plotted are the 4 best beading patterns out of 1600 generated according to the regression model prediction. The objective function is $\qoi$.}
\end{figure}

\begin{figure}[htb]
\centering
\vspace{-0.5in}
\includegraphics[width=\textwidth]{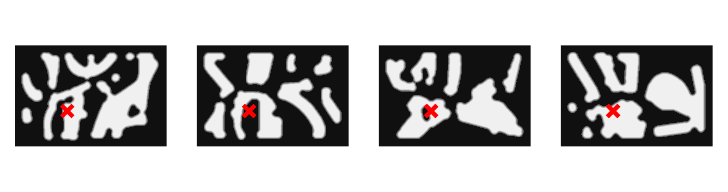}
\includegraphics[width=5.38in]{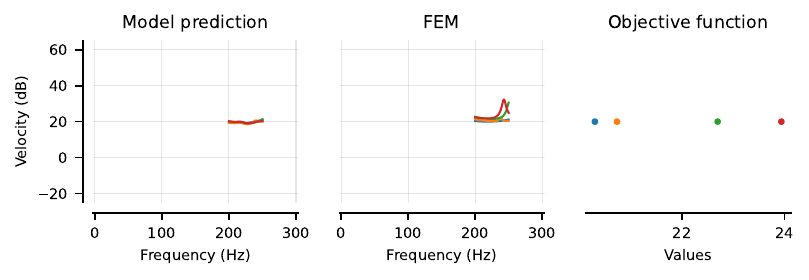}
\caption{Minimizing the response in 200 - 250 Hz. Plate with free rotation at boundary. Plotted are the 4 best beading patterns out of 1600 generated according to the regression model prediction. The objective function is $\qoi$.}
\end{figure}

\begin{figure}[htb]
\centering
\vspace{-0.5in}
\includegraphics[width=\textwidth]{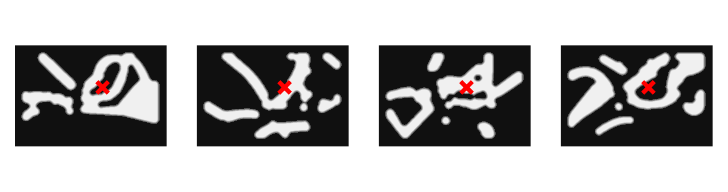}
\includegraphics[width=5.38in]{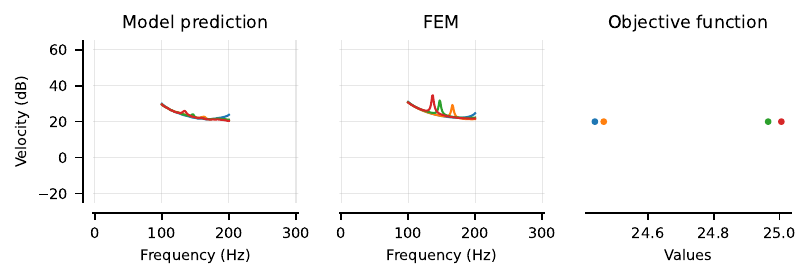}
\caption{Minimizing the response in 100 - 200 Hz. Plate with clamped edges. Plotted are the 4 best beading patterns out of 1600 generated according to the regression model prediction. The objective function is $\qoi$.}
\end{figure}

\begin{figure}[htb]
\centering
\vspace{-0.5in}
\includegraphics[width=\textwidth]{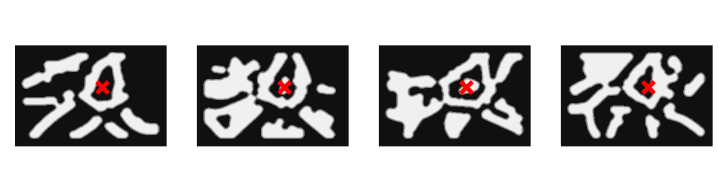}
\includegraphics[width=5.38in]{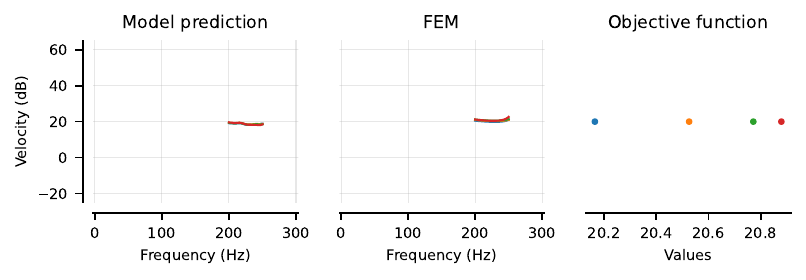}
\caption{Minimizing the response in 200 - 250 Hz. Plate with clamped edges. Plotted are the 4 best beading patterns out of 1600 generated according to the regression model prediction. The objective function is $\qoi$.}
\end{figure}

\begin{figure}[htb]
\centering
\vspace{-0.5in}
\includegraphics[width=\textwidth]{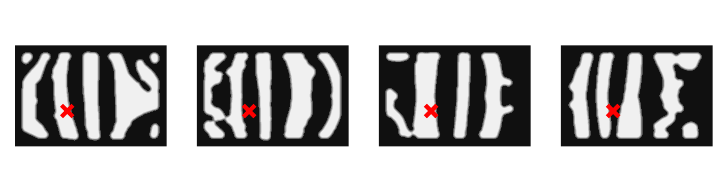}
\includegraphics[width=5.38in]{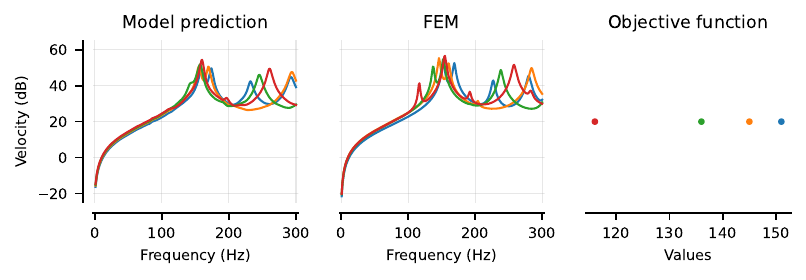}
\caption{Maximizing the first eigenfrequency. Plate with free rotation at boundary. Plotted are the 4 best beading patterns out of 640 generated according to the regression model prediction. The objective function is the value of the first eigenfrequency.}
\end{figure}

\begin{figure}[htb]
\centering
\vspace{-0.5in}
\includegraphics[width=\textwidth]{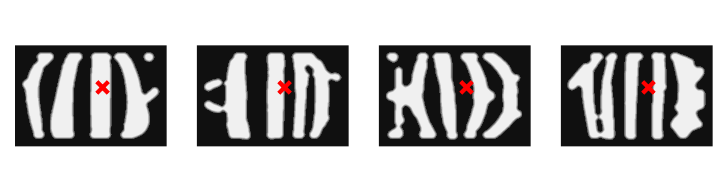}
\includegraphics[width=5.38in]{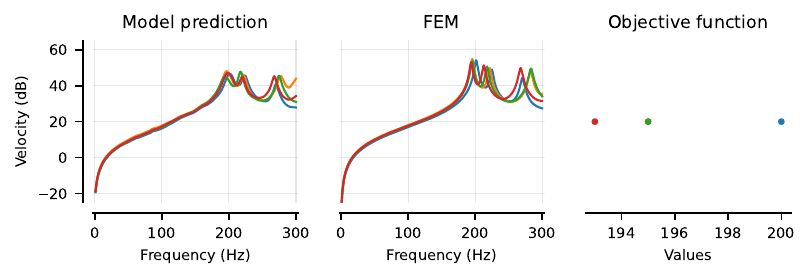}
\caption{Maximizing the first eigenfrequency. Plate with clamped edges. Plotted are the 4 best beading patterns out of 640 generated according to the regression model prediction. The objective function is the value of the first eigenfrequency.}
\end{figure}
\end{document}